\documentclass[prd,onecolumn,nopacs,nofootinbib]{revtex4}

\usepackage{amsmath}
\usepackage{graphicx}
\usepackage{amsfonts}
\usepackage{latexsym}
\usepackage{bbold}
\usepackage{wasysym}
\usepackage{calligra}
\usepackage{float}
\usepackage{ulem}
\usepackage{inputenc}
\usepackage{xspace}
\usepackage{url}
\usepackage{epstopdf}
\usepackage{tikz}
\usepackage{amssymb}
\usepackage{enumitem}

\usepackage{appendix}

\usepackage{amsmath}
\usepackage{amssymb}
\usepackage{graphicx}
\usepackage{amsfonts}
\usepackage{latexsym}
\usepackage{bbold}
\usepackage{wasysym}
\usepackage{calligra}
\usepackage{float}
\usepackage{ulem}
\usepackage{inputenc}
\usepackage{xspace}
\usepackage{url}
\usepackage{epstopdf}
\usepackage{tikz}
\usepackage{amsthm}
\usepackage{soul}

\newcommand{\be}{\begin{equation}}
\newcommand{\ee}{\end{equation}}
\newcommand{\beq} {\begin{equation}}
\newcommand{\eeq} {\end{equation}}
\newcommand{\ba}{\begin{eqnarray}}
\newcommand{\ea}{\end{eqnarray}}

\begin{document}

	\title{	Extended Metric-Affine $f(R)$ Gravity with 
 Dynamical 
 Connection in Vacuum}
	\author{Damianos Iosifidis}
	\affiliation{Laboratory of Theoretical Physics, Institute of Physics, University of Tartu, W. Ostwaldi 1, 50411 Tartu, Estonia.}
	\email{damianos.iosifidis@ut.ee}

	\date{\today}
	\begin{abstract}
We extend the usual vacuum Metric-Affine $f(R)$ Gravity by supplementing it with all parity even quadratic invariants in torsion and non-metricity. As we show explicitly this supplementation drastically changes the status of the Theory which now propagates an additional scalar degree of freedom on top of the graviton. This scalar degree of freedom has a geometric origin as it relates to spacetime torsion and non-metricity. The resulting Theory can be written equivalently as a metric and torsionless Scalar-Tensor Theory whose potential and kinetic term coupling depend on the choice of the function $f(R)$ and the dimensionless parameters of the quadratic invariants respectively.
		
	\end{abstract}
	
	\maketitle
	
	\allowdisplaybreaks
	
	

\section{Introduction}

Geometric extensions of the Riemannian geometry of General Relativity (GR), offer an elegant and effective way to introduce new degrees of freedom apart from the graviton, which are associated to the extra geometric structure.  A wide range of Gravity Theories with such geometric modifications can be formulated in the Metric-Affine Gravity (MAG)\cite{hehl1995metric} formalism. In MAG apart from curvature one has the notions of torsion and non-metricity, both having direct geometrical meanings. The former is responsible for the breaking of  infinitesimal parallelograms when one transports two vectors, one in the direction of the other. The latter quantities how much the lengths of vectors and inner products change when vectors are transported in space.

In MAG the affine-connection (or more appropriately called the $'linear 
 $ $connection'$) is a priory totally independent of the metric and by varying with respect to both, one obtains two separate sets of field equations that need to be solved \cite{Hehl:1976my,hehl1981metric}. By imposing certain conditions on any of the curvature, torsion and non-metricity, certain subcases are extracted. For instance if we impose vanishing torsion and non-metricity and allow only for curvature we obtain the familiar Riemannian geometry and the corresponding metric Theories of Gravity. Imposing vanishing curvature only one arrives at general parallelism \cite{BeltranJimenez:2019odq} and further imposing either vanishing non-metricity or vanishing torsion one is lead to metric teleparallel \cite{aldrovandi2012teleparallel} and symmetric teleparallel \cite{nester1998symmetric,jimenez2018teleparallel} Theories respectively\footnote{For a recent review on the various developments of modified/extended Theories of Gravity, see \cite{CANTATA:2021asi}. }. If only vanishing non-metricity is imposed we arrive to the Riemann-Cartan geometry with Einstein-Cartan and Poincare (gauge) gravity Theories as examples. If only vanishing torsion is assumed we then have a Riemann-Weyl geometry with Einstein-Weyl  Theories an so on.  As already mentioned, in the generic MAG formulation  no constraint is imposed on the aforementioned geometric objects and the underlying manifold admits  both torsion and non-metricity as well as curvature.

In the realm of Riemannian geometry a popular modification of GR consists of extending the Einstein-Hilbert Lagrangian to $f(R)$. In the metric formalism (i.e. vanishing torsion and non-metricity from the onset), it is well known that vacuum $f(R)$ Theories propagate an additional scalar degree of freedom, the so-called scalaron. The resulting Theory is then equivalent to a  generalized Brans-Dicke (BD) Theory with BD parameter $\omega_{0}=0$. On the other hand, vacuum Metric-Affine $f(R)$ Theories behave quite differently, since these can be shown to correspond to GR with Cosmological constant(s), the value(s) of which depend on the solution of some algebraic equation \cite{Ferraris:1992dx}. There is no new dynamical degree of freedom in this case\footnote{In the full quadratic MAG in all potentials (i.e. quadratic in curvature torsion and non-metricity) there are many new propagating degrees of freedom. The full healthy spectrum of quadratic MAG is currently unknown but in recent years there has been some progress in this direction,  see \cite{Percacci:2020ddy,jimenez2020instabilities,Barker:2024ydb,Karananas:2024qrz}}. The situation changes if one adds matter. If no connection-matter couplings are assumed (i.e. for Palatini Theories \cite{olmo2011palatini,Vitagliano:2010pq}\footnote{For some recent developments in Palatini gravity, see for instance \cite{sotiriou2006constraining,Koivisto:2005yc,capozziello2006f,Kubota:2020ehu,demir2022geometric,Gialamas:2023flv,Jarv:2024krk,Capozziello:2024ijv} and for the more general MAG, see \cite{iosifidis2019metric,Iosifidis:2020gth,Iosifidis:2021bad,Baldazzi:2021kaf,Bahamonde:2020fnq,Bahamonde:2022meb,Iosifidis:2023pvz,Rigouzzo:2022yan,Aoki:2023sum,Ducobu:2024grm,Condeescu:2024cjh}. Of course these lists are not exhaustive but rather indicative.}) then hypermomentum is vanishing and the resulting Theory can be shown to correspond to a metric and torsionless Brans-Dicke (BD)Theory with Brans-Dicke parameter $\omega_{0}=-3/2$, see \cite{Sotiriou:2008rp,Sotiriou:2009xt}.  If connection-matter couplings are allowed, that is for non-vanishing hypermomentum, not much can be said since the result is highly sensitive to the choice of these couplings.

A natural question to ask then is whether it is possible to obtain dynamics for vacuum Metric-Affine $f(R)$ Theories by minimally modifying the gravity sector. The most natural choice is of course the inclusion of quadratic scalars built from torsion and non-metricity as well as their mixing. Indeed, already at the level of $f(R)=R$, which corresponds to an extension of Einstein-Cartan Gravity with the incorporation of non-metricity, when matter with hypermomentum is added one encounters a problem coming from the projective invariance of the scalar curvature $R$ which forces the dilation current of hypermomentum to be vanishing \cite{hehl1981metric}. A remedy comes if we add quadratic torsion and non-metricity invariants to the gravity action, which  by explicitly  breaking the projective invariance allow for a non-vanishing dilation current. The addition of these invariants is also motivated by an Effective Field Theory point of view, since their dimensions are the same as those of scalar curvature R (namely $[L^{-2}]$) and therefore there is no principle that excludes their presence. In the current work we add these quadratic terms to the vacuum $f(R)$ Theory and explicitly show that the breaking of projective invariance in this case gives rise to an additional scalar degree of freedom. The resulting Theory is then on-shell equivalent to some metric and torsionless  Scalar-Tensor Theory developed in a Riemannian background. The potential of the Theory depends on the choice of $f(R)$ whereas the kinetic coupling function depends solely on the dimensionless coefficients of the quadratic invariants.

	\section{The setup}
We shall  consider a generic $n-dim$  manifold endowed with a metric $g$ and an independent affine connection $\nabla$, which we denote as ($\mathcal{M}$, g, $\nabla$). The definitions/conventions will be the same with \cite{Iosifidis:2020gth} so for more details the reader may consult the latter. On this non-Riemannian manifold endowed with the metric $g_{\mu\nu}$ and an independent affine connection with components $\Gamma^{\lambda}_{\;\;\;\mu\nu}$, we define the curvature, torsion and non-metricity tensors according to
\beq
R^{\mu}_{\;\;\;\nu\alpha\beta}:= 2\partial_{[\alpha}\Gamma^{\mu}_{\;\;\;|\nu|\beta]}+2\Gamma^{\mu}_{\;\;\;\rho[\alpha}\Gamma^{\rho}_{\;\;\;|\nu|\beta]} \label{R}
\eeq
\beq
S_{\mu\nu}^{\;\;\;\lambda}:=\Gamma^{\lambda}_{\;\;\;[\mu\nu]}
\eeq
\beq
Q_{\alpha\mu\nu}:=- \nabla_{\alpha}g_{\mu\nu}
\eeq
Now, the departure of the affine connection $\Gamma^{\lambda}_{\;\;\;\mu\nu}$ from the Levi-Civita one defines the so-called distortion tensor \cite{schouten1954ricci,hehl1995metric}
\begin{gather}
N^{\lambda}_{\;\;\;\;\mu\nu}:=\Gamma^{\lambda}_{\;\;\;\mu\nu}-\widetilde{\Gamma}^{\lambda}_{\;\;\;\mu\nu}=
\frac{1}{2}g^{\alpha\lambda}(Q_{\mu\nu\alpha}+Q_{\nu\alpha\mu}-Q_{\alpha\mu\nu}) -g^{\alpha\lambda}(S_{\alpha\mu\nu}+S_{\alpha\nu\mu}-S_{\mu\nu\alpha}) \label{N}
\end{gather}
where $\widetilde{\Gamma}^{\lambda}_{\;\;\;\mu\nu}$ is the usual Levi-Civita connection computed solely by the metric and its first derivatives. Once the distortion is given, torsion and non-metricity are readily obtained (see for instance \cite{iosifidis2019metric})
\beq
S_{\mu\nu\alpha}=N_{\alpha[\mu\nu]}\;\;,\;\;\; Q_{\nu\alpha\mu}=2 N_{(\alpha\mu)\nu} \label{QNSN}
\eeq

Out of torsion we can construct a vector as well as a pseudo-vector. Our definitions for the torsion vector and pseudo-vector are
\beq
S_{\mu}:=S_{\mu\lambda}^{\;\;\;\;\lambda} \;\;, \;\;\;
t_{\mu}:=\epsilon_{\mu\alpha\beta\gamma}S^{\alpha\beta\gamma} 
\eeq
respectively. Note that the former is defined for any dimension while the latter only for $n=4$. In the forthcoming discussion we will only need the former. Continuing with non-metricity, we define the Weyl  and the second non-metricity vector according to
\beq
Q_{\alpha}:=Q_{\alpha\mu\nu}g^{\mu\nu}\;,\;\; q_{\nu}=Q_{\alpha\mu\nu}g^{\alpha\mu}
\eeq
From the curvature tensor we can construct the three contractions
\beq
R_{\nu\beta}:=R^{\mu}_{\;\;\nu\mu\beta}	
\eeq
\beq
\hat{R}_{\alpha\beta}:=R^{\mu}_{\;\;\mu\alpha\beta}	
\eeq
\beq
\breve{R}^{\mu}_{\;\;\beta}:=R^{\mu}_{\;\;\nu\alpha\beta}	g^{\nu\alpha}
\eeq
The first one is generalized Ricci tensor (which is not symmetric in general), the second goes by the name homothetic curvature and the last one is the so-called co-Ricci tensor. Observe that in constructing the Ricci scalar and homothetic curvature no metric is required, whereas in order to define the co-Ricci a metric should be given. Let us mention that the generalized scalar curvature is still uniquely defined, since
\beq
R:=g^{\mu\nu}R_{\mu\nu}=-g^{\mu\nu}\breve{R}_{\mu\nu} \;\; ,\;\;\; g^{\mu\nu}\hat{R}_{\mu\nu}=0
\eeq

Furthermore, let us mention that by virtue of ($\ref{N}$) each quantity can be split into its Riemannian part (i.e. computed with respect to the Levi-Civita connection) plus non-Riemannian contributions. For instance inserting the connection decomposition ($\ref{N}$) into the definition ($\ref{R}$) we obtain for the curvature tensor \footnote{Quantities with\;  $\widetilde{}$\;  will always denote Riemannian parts  unless otherwise stated.} 
\beq
{R^\mu}_{\nu \alpha \beta} = \widetilde{R}^\mu_{\phantom{\mu} \nu \alpha \beta} + 2 \widetilde{\nabla}_{[\alpha} {N^\mu}_{|\nu|\beta]} + 2 {N^\mu}_{\lambda|\alpha} {N^\lambda}_{|\nu|\beta]} \,, \label{decomp}
\eeq
The last decomposition is very useful and we are going to be using it later on  in order to express the metric field equations of our Theory in Einstein-like form plus extra fields. For instance, with the use of the above the post-Riemannian expansion of the Ricci scalar reads
\begin{gather}
R=\tilde{R}+ \frac{1}{4}Q_{\alpha\mu\nu}Q^{\alpha\mu\nu}-\frac{1}{2}Q_{\alpha\mu\nu}Q^{\mu\nu\alpha}    -\frac{1}{4}Q_{\mu}Q^{\mu}+\frac{1}{2}Q_{\mu}q^{\mu}+S_{\mu\nu\alpha}S^{\mu\nu\alpha}-2S_{\mu\nu\alpha}S^{\alpha\mu\nu}-4S_{\mu}S^{\mu} \nonumber \\ +2 Q_{\alpha\mu\nu}S^{\alpha\mu\nu}+2 S_{\mu}(q^{\mu}-Q^{\mu}) +\tilde{\nabla}_{\mu}(q^{\mu}-Q^{\mu}-4S^{\mu})
\end{gather}

Finally let us note that out for the distortion (\ref{N}), one can construct three independent vectors as
	\beq
	N^{(1)}_{\mu}:=N_{\alpha\beta\mu}g^{\alpha\beta}\;\;, \;\; N^{(2)}_{\mu}:=N_{\alpha\mu\beta}g^{\alpha\beta}\;\;, \;\; N^{(3)}_{\mu}:=N_{\mu\alpha\beta}g^{\alpha\beta}
	\eeq
	which we may call the $1^{st}$, $2^{nd}$ and $3^{rd}$ contractions of $N_{\alpha\mu\nu}$. Let us also observe that these three vectors are related to the torsion and non-metricity vectors defined above. Indeed, starting from the relations \eqref{QNSN} and taking the three independent contractions, recalling also the definitions of torsion and non-metricity vectors, it follows that
\beq	N_{(1)\mu}:=N^{\lambda}_{\;\;\lambda\mu}=\frac{1}{2}Q_{\mu} \label{NSQ1}
	\eeq
	\beq
	N_{(2)\mu}:=N^{\lambda}_{\;\;\mu\lambda}=\frac{1}{2}Q_{\mu}+2 S_{\mu}
	\eeq
	\beq
	N_{(3)\mu}=N_{\mu\alpha\beta}g^{\alpha\beta}=q_{\mu}-\frac{1}{2}Q_{\mu}-2 S_{\mu} \label{NSQ3}
	\eeq
Having defined  the necessary geometric  tools we may now focus on the applications. We shall start with a simple model in order to get some intuition on the procedure and then attack the general case.

\section{A simple Theory}
Let us start with the simple  of Theory
\beq
S=\frac{1}{2 \kappa}\int d^{n}x \sqrt{-g}\Big( f(R) +\beta Q_{\mu}Q^{\mu}+\gamma S_{\mu}S^{\mu} \Big) \label{Thesimp}
\eeq
where $R$ is the generalized scalar curvature and $\beta$ and $\gamma$ are dimensionless constants that are both non-zero. If only one of them is present, it can be easily shown, by employing the field equations, that the resulting Theory is identical to GR with a Cosmological constant, just like in the usual vacuum Metric-Affine $f(R)$. Therefore it is essential to keep both of them non-vanishing in order to establish non-trivial dynamics, as we show below.
The above Theory is simple enough to explicitly work out the calculations in full detail and yet  general enough to feature the main characteristics of the more complicated full Theory. Variations with respect to the metric and the connection give the field equations
\begin{gather}
    f' R_{(\mu\nu)}-\frac{f}{2}g_{\mu\nu}+\gamma\Big( S_{\mu}S_{\nu}-\frac{1}{2}S_{\alpha}S^{\alpha}g_{\mu\nu} \Big)+\beta\Big( Q_{\mu}Q_{\nu}-\frac{1}{2}Q_{\alpha}Q^{\alpha}g_{\mu\nu} \Big)-2 \beta g_{\mu\nu}\frac{1}{\sqrt{-g}}\partial_{\alpha}(\sqrt{-g}Q^{\alpha})=0 \label{mfe}
\end{gather}
\beq
f' P_{\lambda}{}^{\mu\nu}+\delta_{\lambda}^{\nu}\partial_{\mu}f'-g^{\mu\nu}\partial_{\lambda}f' +4 \beta \delta_{\lambda}^{\mu}Q^{\nu}+2 \gamma S^{[\mu}\delta_{\lambda}^{\nu]}=0 \label{confe}
\eeq
respectively, where
\begin{gather}
	P_{\lambda}^{\;\;\;\mu\nu}=-\frac{\nabla_{\lambda}(\sqrt{-g}g^{\mu\nu})}{\sqrt{-g}}+\frac{\nabla_{\sigma}(\sqrt{-g}g^{\mu\sigma})\delta^{\nu}_{\lambda}}{\sqrt{-g}} 
	+2(S_{\lambda}g^{\mu\nu}-S^{\mu}\delta_{\lambda}^{\nu}+g^{\mu\sigma}S_{\sigma\lambda}^{\;\;\;\;\nu})  \nonumber \\
=\delta^{\nu}_{\lambda}\left( \tilde{Q}^{\mu}-\frac{1}{2}Q^{\mu}-2 S^{\mu} \right) + g^{\mu\nu}\left( \frac{1}{2}Q_{\lambda}+2 S_{\lambda} \right)-( Q_{\lambda}^{\;\;\;\mu\nu}+2 S_{\lambda}^{\;\;\;\;\mu\nu})
	\end{gather}
is the so-called Palatini tensor.
Let us firstly focus on the connection field equations (\ref{confe}). Tracing in $\lambda=\mu$ it follows that
\beq
Q_{\mu}=\frac{(n-1)}{n}\frac{\gamma}{4 \beta}S_{\mu} \label{Qs}
\eeq
In addition, taking the other two independent traces\footnote{Namely contracting one time with $\delta^{\lambda}_{\nu}$ and another with $g_{\mu\nu}$.} and adding them up, using also the latter, we deduce
\beq
q_{\mu}=\frac{(n-1)}{n^{2}}\gamma \left[ \frac{1}{4 \beta}-\frac{(n+2)}{f'} \right] S_{\mu}
\eeq
By subtracting the resulting trace equations and using both of the above we find
\beq
S_{\mu}=\frac{\partial_{\mu}f'}{A f' +B} \label{Sf}
\eeq
where
\beq
A=\frac{(n-2)\gamma}{2}\left[- \frac{(n-1)}{4 \beta n^{2}}+\frac{4}{\gamma (n-1)} \right]\;\;, \;\;\;\; B=-\frac{2 \gamma}{n^{2}}
\eeq
Equations (\ref{Qs})-(\ref{Sf}) together imply that all torsion and non-metricity vectors are proportional to one another and sourced by $f'$. Further set\footnote{Of course here the $\widetilde{}$ does not correspond to some Riemannian quantity but is merely a  symbol used for the redefinition of constants.}
\beq
\lambda=\frac{(n-1)}{n}\frac{\gamma}{4 \beta} \;\;, \;\; \widetilde{A}=\frac{\lambda}{n}=-\frac{(n-1)}{8 \beta}B\;\;, \;\; \widetilde{B}=-\frac{4 \beta \lambda (n+2)}{n}
\eeq
then
\beq
Q_{\mu}=\lambda S_{\mu}=\lambda \frac{\partial_{\mu}f'}{A f' +B}=\frac{\lambda}{A}\partial_{\mu}\Big[\ln{(A f'+B)}\Big]
\eeq
and
\beq
q_{\mu}=\Big[ \widetilde{A}+\frac{\widetilde{B}}{f'} \Big]S_{\mu}=\partial_{\mu}\left[ \frac{\widetilde{B}}{B}\ln{(A f'+ B)}+\frac{\widetilde{A}}{A}\ln{\Big( \frac{f'}{A f'+B }\Big)}\right]
\eeq
Note that all the torsion and non-metricity vectors are exact. Substituting everything back to the connection field equations, and also setting
\beq
\psi_{\nu}:=\frac{\partial_{\nu} f'}{f'}\;\;, \;\; F:=\frac{\gamma}{(Af'+B)}
\eeq
we solve for the Palatini tensor entirely in  terms of $f'$ and its first derivatives:
\beq
P_{\alpha\mu\nu}=F\left( \frac{1}{n}g_{\alpha\mu}\psi_{\nu}-g_{\alpha\nu}\psi_{\mu} \right)+2 \psi_{[\alpha}g_{\mu]\nu}
\eeq
As a good cross-check we  confirm that the latter is obviously traceless in its first two indices as it should. With this result we can then find the exact form of the distortion (see \cite{Iosifidis:2018jwu} for details):
\beq
N_{\alpha\mu\nu}=\Omega_{1}g_{\alpha\mu}\psi_{\nu}+\Omega_{2}g_{\alpha\nu}\psi_{\mu}+\Omega_{3}g_{\mu\nu}\psi_{\alpha} \label{distor}
\eeq
where
\beq
2 \Omega_{1}=\frac{(n+1)}{n}F+\frac{\widetilde{A}f' +\widetilde{B}}{A f'+B}\;, \;\; \Omega_{2}=\frac{1}{(n-2)}\left(1+\frac{F}{n}\right) \;, \;\; \Omega_{3}=-\frac{1}{(n-2)}\left(1+\frac{(n-1)}{n}F\right)
\eeq
Therefore we see that the full torsion and non-metricity are sourced by the derivative of $f'$. Let us now turn our attention to the metric field equations (\ref{mfe}). Taking the trace of the latter and also using the fact that $Q_{\mu}=\lambda S_{\mu}$, it follows 
\beq
f' R-\frac{n}{2}f+\left( 1-\frac{n}{2}\right) (\beta \lambda^{2}+\gamma)S_{\mu}S^{\mu}=2\beta n \frac{1}{\sqrt{-g}}\partial_{\alpha}(\sqrt{-g}Q^{\alpha}) \label{tra}
\eeq
Let us now consider the field redefinition
\beq
\phi:=f'
\eeq
Given that the latter is invertible we can solve it for $R$ and express
\beq
R=\chi(\phi)
\eeq
In addition, setting
\beq
G(\phi)=\phi\chi(\phi)-\frac{n}{2}f(\chi(\phi))
\eeq
and
using all the above the trace equation (\ref{tra}) becomes
\beq
\Box_{g} \phi=\frac{2(A\phi+B)}{(n-1)\gamma}G(\phi)+\left(\frac{n-2}{n-1} \right)\left[ 1+\left(\frac{n-1}{4 n}\right)^{2}\frac{\gamma}{\beta}\right]\frac{(\partial \phi)^{2}}{(A \phi +B)}
\eeq
where $\Box_{g}:=\frac{1}{\sqrt{-g}}\partial_{\mu}(\sqrt{-g}g^{\mu\nu}\partial_{\nu})$ is the usual box operator.  This describes the evolution of a scalar mode. Therefore we  conclude that the Theory (\ref{Thesimp}) propagates an  additional scalar degree of freedom aside the graviton.
Notice that for the parameter choice $\beta =-\frac{(n-1)^{2}}{16 n^{2}}\gamma$ the latter becomes a generalized Klein-Gordon equation with scalar field potential $V'(
\phi)=2 \frac{A\phi +B}{(n-1)\gamma}G(\phi)$. Obviously, in 2-dimensions this holds true always without the need to impose any constraint on the parameters.

Therefore we see that the supplementation of the Metric-Affine $f(R)$ Lagrangian with quadratic torsion and non-metricity invariants has a dramatic affect on the dynamics. Indeed, it is well known that vacuum Metric-Affine $f(R)$ gravity, in contrast to metric $f(R)$, corresponds to GR with cosmological constant(s), the value(s) of which depend on the explicit form of the function $f(R)$, see \cite{Ferraris:1992dx,sotiriou2007metric}. As we showed in detail here, the addition of two quadratic torsion and non-metricity invariants to the $f(R)$ Lagrangian brings out an additional scalar degree of freedom and the original Metric-Affine Theory is actually equivalent to a specific metric and torsionless Scalar-Tensor Theory. Let us elaborate more on this point below by also establishing this equivalence at the level of the action.

\subsection{Equivalent Scalar-Tensor Theory}

Let us now  prove also formally the result of the previous subsection, namely that the Theory (\ref{Thesimp}) propagates an additional scalar degree of freedom, alongside the graviton, by establishing equivalence with a metric and torsionless Scalar-Tensor Theory. We start in the usual manner \cite{OHanlon1972IntermediaterangeG}(see also \cite{Teyssandier:1983zz,Sotiriou:2006hs}), by introducing an auxiliary field $\chi$ and write down the action
\beq
S=\frac{1}{2 \kappa}\int d^{n}x \sqrt{-g}\Big[ f(\chi)+f'(\chi)(R-\chi)+\beta Q_{\mu}Q^{\mu}+\gamma S_{\mu}S^{\mu} \Big] \label{SactQS}
\eeq
where $f'(\chi)=\partial f/\partial \chi$. Of course, varying with respect to $\chi$ one obtains as usual  $\chi=R$ provided that $f''\neq 0$. When this algebraic relation is put back on the above one recovers  equivalence with (\ref{Thesimp}). Considering the field redefinition $\Phi=f'(\chi)$ and further setting $V(\Phi)=\chi(\Phi)\Phi-f(\chi(\Phi))$ the latter recasts to
\beq
S=\frac{1}{2 \kappa}\int d^{n}x \sqrt{-g}\Big[ \Phi R -V(\Phi)+\beta Q_{\mu}Q^{\mu}+\gamma S_{\mu}S^{\mu} \Big]
\eeq
Now, using the the post-Riemannian expansion of the scalar curvature, connection field equations (\ref{confe}), and the definitions of the previous subsection we readily obtain 
\beq
\sqrt{-g} \Phi R=\sqrt{-g}\Phi \tilde{R}+\sqrt{-g}\left(\frac{n-1}{n-2}\right)\frac{(\partial \Phi)^{2}}{\Phi}\left(1-\frac{F^{2}}{n^{2}} \right)
\eeq
Using also eqns (\ref{Qs}) and (\ref{Sf}) the above action is computed to be on-shell equivalent to
\beq
S=\frac{1}{2 \kappa}\int d^{n}x \sqrt{-g}\left[\Phi \tilde{R} 
-V(\Phi)+\left( \frac{n-1}{n-2}\right)\Big( (A \Phi+B)^{2}-\frac{\gamma^{2}}{n^{2}}+\frac{(n-2)}{(n-1)}(\beta \lambda^{2}+\gamma)\Phi \Big)\frac{(\partial \Phi)^2}{\Phi (A \Phi+B)^2}
  \right]
\eeq
 which corresponds to a metric-compatible and torsionless Scalar-Tensor Theory! It is worth mentioning that if we fix $\beta$ and consider a small torsion coupling $\gamma \ll 1$ the Theory becomes a Brans-Dicke (BD) Theory
with BD parameter $\omega_{0}=-(n-1)/(n-2)$. Quite intriguingly the same equivalence\footnote{In particular for $n=4$ which corresponds to our physical world and consequently $\omega_{0}=-3/2$.} holds for Palatini f(R) Gravity \cite{Sotiriou:2006hs} and also Metric-Affine f(R) Gravity with the addition of the Hojman (or Holst) term \cite{Iosifidis:2020dck}. However, the case under consideration here is more general since, as we have shown, the Theory (\ref{Thesimp}) is equivalent to a specific Scalar-Tensor generalization of the Brans-Dicke Theory.

To make this point more precise and offer a direct comparison with the usual Scalar-Tensor action, we define 
\beq
-\omega(\Phi)=\left( \frac{n-1}{n-2}\right)\Big( (A \Phi+B)^{2}-\frac{\gamma^{2}}{n^{2}}+\frac{(n-2)}{(n-1)}(\beta \lambda^{2}+\gamma)\Phi \Big)\frac{1}{ (A \Phi+B)^2}
\eeq
and the above takes the usual Scalar-Tensor form
\beq
S=\frac{1}{2 \kappa}\int d^{n}x \sqrt{-g}\left[\Phi \tilde{R} 
-V(\Phi)-\frac{\omega(\Phi)}{\Phi}(\partial \Phi)^{2}
  \right]
\eeq
which is the conventional form of Scalar-Tensor Theories in the so-called Jordan (or string) frame. Of course by performing a conformal transformation, with the appropriate conformal factor, the $\Phi$-factor multiplying the Ricci scalar can be removed and the Theory is then written in the Einstein frame. 
With the intuition gained by studying this simple model we are now ready to tackle the more general case.

\section{Including all quadratic invariants}

Our previous considerations can be generalized by adding to the $f(R)$ Lagrangian all 11 parity even quadratic invariants in torsion and non-metricity, namely supplementing $f(R)$ with \cite{Iosifidis:2018zwo}
\begin{gather}
\mathcal{L}_{2}=b_{1}S_{\alpha\mu\nu}S^{\alpha\mu\nu} +
	b_{2}S_{\alpha\mu\nu}S^{\mu\nu\alpha} +
	b_{3}S_{\mu}S^{\mu}+
	a_{1}Q_{\alpha\mu\nu}Q^{\alpha\mu\nu} +
	a_{2}Q_{\alpha\mu\nu}Q^{\mu\nu\alpha} \nonumber \\ +
	a_{3}Q_{\mu}Q^{\mu}+
	a_{4}q_{\mu}q^{\mu}+
	a_{5}Q_{\mu}q^{\mu} 
	+c_{1}Q_{\alpha\mu\nu}S^{\alpha\mu\nu}+
	c_{2}Q_{\mu}S^{\mu} +
	c_{3}q_{\mu}S^{\mu} \label{L2}
\end{gather}
The only difference is that now the computations are much more involved, however the take-home message remains the same, namely this MAG Theory is equivalent to some metric Scalar-Tensor Theory.   To see this let us consider the  extension of (\ref{Thesimp}) where all the above quadratic invariants are included. We have
	\begin{gather}
	S
	=\frac{1}{2 \kappa}\int d^{4}x \sqrt{-g} \Big[ f(R)+\
	b_{1}S_{\alpha\mu\nu}S^{\alpha\mu\nu} +
	b_{2}S_{\alpha\mu\nu}S^{\mu\nu\alpha} +
	b_{3}S_{\mu}S^{\mu}+
	a_{1}Q_{\alpha\mu\nu}Q^{\alpha\mu\nu} +
	a_{2}Q_{\alpha\mu\nu}Q^{\mu\nu\alpha} \nonumber \\ +
	a_{3}Q_{\mu}Q^{\mu}+
	a_{4}q_{\mu}q^{\mu}+
	a_{5}Q_{\mu}q^{\mu} 
	+c_{1}Q_{\alpha\mu\nu}S^{\alpha\mu\nu}+
	c_{2}Q_{\mu}S^{\mu} +
	c_{3}q_{\mu}S^{\mu}
	\Big]  \label{S}
	\end{gather}
Varying with respect to the connection, it follows that
	\begin{gather}
f'\left( \frac{Q_{\lambda}}{2}+2 S_{\lambda}\right)g^{\mu\nu}-f' (Q_{\lambda}^{\;\;\mu\nu}+2 S_{\lambda}^{\;\;\mu\nu})+f' \left( q^{\mu} -\frac{Q^{\mu}}{2}-2 S^{\mu}\right)\delta_{\lambda}^{\nu} \nonumber \\+\delta_{\lambda}^{\nu}\partial_{\mu}f'-g^{\mu\nu}\partial_{\lambda}f'+4 a_{1}Q^{\nu\mu}_{\;\;\;\;\lambda}+2 a_{2}(Q^{\mu\nu}_{\;\;\;\;\lambda}+Q_{\lambda}^{\;\;\;\mu\nu})+2 b_{1}S^{\mu\nu}_{\;\;\;\;\lambda}+2 b_{2}S_{\lambda}^{\;\;\;[\mu\nu]} \nonumber \\
+c_{1}\Big( S^{\nu\mu}_{\;\;\;\;\lambda}-S_{\lambda}^{\;\;\;\nu\mu}+Q^{[\mu\nu]}_{\;\;\;\;\;\lambda}\Big)+\delta_{\lambda}^{\mu}\Big( 4 a_{3}Q^{\nu}+2 a_{5}q^{\nu}+2 c_{2}S^{\nu}\Big)+\delta_{\lambda}^{\nu}\Big(  a_{5}Q^{\mu}+2 a_{4}q^{\mu}+ c_{3}S^{\mu}\Big) \nonumber \\
+g^{\mu\nu}\Big(a_{5} Q_{\lambda}+2 a_{4}q_{\lambda}+c_{3}S_{\lambda} \Big)+\Big( c_{2} Q^{[\mu}+ c_{3}q^{[\mu}+2 b_{3}S^{[\mu}\Big) \delta^{\nu]}_{\lambda} =0 \label{Gfieldeqs}
\end{gather}
Taking the 3 independent contractions of the latter, we obtain the system
\begin{gather}
    a_{11}Q_{\mu}+a_{12}q_{\mu}+a_{13}S_{\mu}=0 \\ \label{a11}
    (b_{21}f'+c_{21})Q_{\mu}+(b_{22}f' +c_{22})q_{\mu}+(b_{23}f'+c_{23})S_{\mu}=-(n-1)\partial_{\mu}f'  \\
     (b_{31}f'+c_{31})Q_{\mu}+(b_{32}f' +c_{31})q_{\mu}+(b_{33}f'+c_{33})S_{\mu}=(n-1)\partial_{\mu}f' \label{a12}
\end{gather}
where $a_{ij}, b_{ij}$ and $c_{ij}$ are some long linear combinations of the original coefficients $a_{i}, b_{j}, c_{k}$, which we list in the appendix A.  Further setting $a_{ij}=b_{ij}f' +c_{ij}$ for $i\neq 1$ and defining the matrix $\mathcal{A}=\{ a_{ij}\}$ we can express the above system in matrix form as
\beq
\mathcal{A}X_{\mu}=Y_{\mu} \label{Aeq}
\eeq
with $X_{\mu}=(Q_{\mu},q_{\mu},S_{\mu})^T$ and $Y_{\mu}=(n-1)(\partial_{\mu}f')(0,-1,1)^T$. The solutions of (\ref{Aeq}) depend on the properties of the matrix $\mathcal{A}$. On the reasonable assumption that the latter is non-singular (i.e. $det(\mathcal{A})\neq 0$) one can of course formally write down the non-degenerate solution as
\beq
X_{\mu}=\mathcal{A}^{-1}Y_{\mu} \label{XAY}
\eeq
where $\mathcal{A}^{-1}$ is the inverse of $\mathcal{A}$. Let $\{ \tilde{a}_{ij}\}$ be the elements of the inverse matrix $\mathcal{A}^{-1}$. Then,  thanks to the fact that the $a_{1i}'s$ do not depend on $f'$, the $\tilde{a}_{ij}'s$ for $j \neq 1$ have the form
\beq
\tilde{a}_{ij}=\frac{\tilde{A}_{ij}f'+\tilde{B}_{ij}}{A_{0}(f')^{2}+B_{0}f'+C_{0}}=\frac{\tilde{A}_{ij}f'+\tilde{B}_{ij}}{P_{2}(f')} \;\;\;, \;\;\; j \neq 1
\eeq
with the precise form of these elements, and the various constants appearing, given in the appendix A. Note that the denominator
\beq
P_{2}(f'):=A_{0}(f')^{2}+B_{0}f'+C_{0}
\eeq
is universal for all elements!\footnote{Of course this comes from the determinant of $\mathcal{A}$ involved when computing the inverse.} 
The above matrix elements are special, with numerator being linear in $f'$, in contrast to the $\tilde{a}_{i1}'s$ whose numerator is also quadratic in $f'$. However, since $Y_{\mu}\propto (0,-1,1)^T$ the latter components are irrelevant for the actual form of the torsion and non-metricity vectors. Indeed, equating the components of the left and right hand sides of (\ref{XAY}) it follows that
\beq
Q_\mu=(n-1)(\tilde{a}_{13}-\tilde{a}_{12})\partial_{\mu}f'=\frac{A_{1}f'+B_{1}}{P_{2}(f')}\partial_{\mu}f' \label{Q}
\eeq
\beq
q_\mu=(n-1)(\tilde{a}_{23}-\tilde{a}_{22})\partial_{\mu}f'=\frac{A_{2}f'+B_{2}}{P_{2}(f')}\partial_{\mu}f'
\eeq
\beq
S_\mu=(n-1)(\tilde{a}_{33}-\tilde{a}_{32})\partial_{\mu}f'=\frac{A_{3}f'+B_{3}}{P_{2}(f')}\partial_{\mu}f' \label{Smu}
\eeq
where we have set $A_{i}=(n-1)(\tilde{A}_{i3}-\tilde{A}_{i2})$ and $B_{i}=(n-1)(\tilde{B}_{i3}-\tilde{B}_{i2})$. Note that all of these vectors are exact. Now, after substituting the latter forms of torsion and non-metricity vectors back in (\ref{confe}) and using the result of \cite{iosifidis2019exactly} (or more generally \cite{Iosifidis:2021ili}) one can obtain the exact form for the distortion. Alternatively, and more quickly, given that the right-hand side of (\ref{confe}) would consist only of the metric tensor and the derivative of $f'$, the distortion tensor has an expression formally identical to (\ref{distor}), viz.
\beq
N_{\alpha\mu\nu}=\Omega_{1}g_{\alpha\mu}\psi_{\nu}+\Omega_{2}g_{\alpha\nu}\psi_{\mu}+\Omega_{3}g_{\mu\nu}\psi_{\alpha} \label{distor2}
\eeq
with the only difference being that now the coefficient functions $\Omega_{i}$, $i=1,2,3$ will be slightly more involved. Indeed, using the connecting equations (\ref{NSQ1})-(\ref{NSQ3}) between the distortion vectors and those of torsion and non-metricity, given also  (\ref{Q})-(\ref{Smu}), it follows that 
\beq
N_{\mu}^{(i)}=\frac{A_{i+3}f'+B_{i+3}}{P_{2}(f')}\partial_{\mu}f'\;, \;\; i=1,2,3
\eeq
where
\beq
A_{4}=A_{1}/4 \;\;,\;\; A_{5}=X_{1}/2+2 A_{3}\;\;, \;\; A_{6}=A_{2}-X_{1}/2 -2 A_{3}
\eeq
and same for the $B_{i}$'s. Then, taking the trace of (\ref{distor2}) and using the above forms of the distortion traces we readily find the $\Omega$ functions,
\beq
\Omega_{i}=\frac{f'}{P_{2}(f')}(C_{i}f'+D_{i}) \;\;, i=1,2,3 \label{omega123}
\eeq
where 
\beq
C_{1}=\frac{1}{(n-1)(n+2)}\Big( (n+1)A_{4}-A_{5}-A_{6}\Big)
\eeq
\beq
C_{2}=\frac{1}{(n-1)(n+2)}\Big( -A_{4}+(n+1)A_{5}-A_{6}\Big)
\eeq
\beq
C_{3}=\frac{1}{(n-1)(n+2)}\Big( -A_{4}-A_{5}+(n+1)A_{6}\Big)
\eeq
and exactly the same for the $D_{i}$'s with the mere replacement of C by D and A replaced by B in the above. Let us note that all $\Omega_{i}$'s are given by ratios of quadratic polynomials in $f'$. 

We have now everything we need in order to reveal the extra scalar mode as promised. To this end, we now vary (\ref{S}) with respect to the metric, to derive the metric field equations
	\beq
f' R_{(\mu\nu)}-\frac{f}{2}g_{\mu\nu}-\frac{\mathcal{L}_{2}}{2}g_{\mu\nu}+\frac{1}{\sqrt{-g}}(2S_{\alpha}-\nabla_{\alpha})\Big( \sqrt{-g}(W^{\alpha}_{\;\;\;(\mu\nu)}+\Pi^{\alpha}_{\;\;\;(\mu\nu)})\Big) +A_{(\mu\nu)}+B_{(\mu\nu)}+C_{(\mu\nu)}=\kappa T_{\mu\nu} \label{metricf}
\eeq
where 
\beq
W^{\alpha}_{\;\;(\mu\nu)}=2 a_{1}Q^{\alpha}_{\;\;\mu\nu}+2 a_{2}Q_{(\mu\nu)}^{\;\;\;\;\alpha}+(2 a_{3}Q^{\alpha}+a_{5}q^{\alpha})g_{\mu\nu}+(2 a_{4}q_{(\mu} + a_{5}Q_{(\mu})\delta^{\alpha}_{\nu)}
\eeq
\beq
\Pi^{\alpha\mu\nu} = c_{1}S^{\alpha\mu\nu}+c_{2}g^{\mu\nu}S^{\alpha}+c_{3}g^{\alpha\mu}S^{\nu}
\eeq

\beq
A_{\mu\nu}=a_{1}(Q_{\mu\alpha\beta}Q_{\nu}^{\;\;\alpha\beta}-2 Q_{\alpha\beta\mu}Q^{\alpha\beta}_{\;\;\;\;\nu})-a_{2}Q_{\alpha\beta(\mu}Q^{\beta\alpha}_{\;\;\;\;\nu)}
+a_{3}(Q_{\mu}Q_{\nu}-2 Q^{\alpha}Q_{\alpha\mu\nu})-a_{4}q_{\mu}q_{\nu}-a_{5}q^{\alpha}Q_{\alpha\mu\nu}
\eeq	
\beq
B_{\mu\nu}=b_{1}(2S_{\nu\alpha\beta}S_{\mu}^{\;\;\;\alpha\beta}-S_{\alpha\beta\mu}S^{\alpha\beta}_{\;\;\;\;\nu})-b_{2}S_{\nu\alpha\beta}S_{\mu}^{\;\;\;\beta\alpha}+b_{3}S_{\mu}S_{\nu} 
\eeq
\beq
C_{\mu\nu}=\Pi_{\mu\alpha\beta}Q_{\nu}^{\;\;\;\alpha\beta}	-( c_{1}S_{\alpha\beta\nu}Q^{\alpha\beta}_{\;\;\;\;\mu}+c_{2}S^{\alpha}Q_{\alpha\mu\nu}+c_{3}S^{\alpha}Q_{\mu\nu\alpha})=c_{1}(Q_{\mu}^{\;\;\;\alpha\beta}S_{\nu\alpha\beta}-S_{\alpha\beta\mu}Q^{\alpha\beta}_{\;\;\;\;\nu})+c_{2}(S_{\mu}Q_{\nu}-S^{\alpha}Q_{\alpha\mu\nu})
\eeq
Taking the trace of (\ref{metricf}), it follows that
	\beq
f' R-\frac{n}{2}f+\Big( 1-\frac{n}{2} \Big) \mathcal{L}_{2}-\frac{1}{\sqrt{-g}}\tilde{\nabla}_{\alpha}\Big(\sqrt{-g}(\Pi^{\alpha}+W^{\alpha})\Big)=0 \label{trace}
\eeq
where
\beq
\Pi^{\alpha}:=\Pi^{\alpha}_{\;\;\mu\nu}g^{\mu\nu}=(c_{1}+n c_{2}+c_{3})S^{\alpha}
\eeq
\beq
W^{\alpha}:=W^{\alpha}_{\;\;\mu\nu}g^{\mu\nu}=(2 a_{1}+2 n a_{3}+a_{5})Q^{\alpha}+(2 a_{2} +2 a_{4}+n a_{5})q^{\alpha}
\eeq
Given the forms of torsion and non-metricity vectors and the above relations we compute
\beq
\Pi^{\alpha}+W^{\alpha}=\frac{(\lambda_{1}f'+\lambda_{2})}{P_{2}(f')}\partial^{\alpha}f'
\eeq
with
\beq
\lambda_{1}=(2 a_{1}+2 n a_{3}+a_{5})A_{1}+(2 a_{2} +2 a_{4}+n a_{5}) A_{2}+  (c_{1}+n c_{2}+c_{3})A_{3}
\eeq
\beq
\lambda_{2}=(2 a_{1}+2 n a_{3}+a_{5})B_{1}+(2 a_{2} +2 a_{4}+n a_{5}) B_{2}+  (c_{1}+n c_{2}+c_{3})B_{3}
\eeq
In addition, from (\ref{L2}) and with (\ref{distor2}) at hand we easily find
\beq
\mathcal{L}_{2}=(\beta_{11}\Omega_{1}^{2}+\beta_{22}\Omega_{2}^{2}+\beta_{33}\Omega_{3}^{2}+2\beta_{12}\Omega_{1}\Omega_{2}+2\beta_{23}\Omega_{2}\Omega_{3}+2\beta_{31}\Omega_{3}\Omega_{1})\frac{(\partial \phi)^{2}}{\phi^2}
\eeq
the expressions for the coefficients are given in the appendix B. Finally, after the field redefinition $f'(R)=\phi\Rightarrow  R=\chi(\phi)$ and with the abbreviations 
\beq
G(\phi)=\phi \chi(\phi)-\frac{n}{2}f(\chi(\phi))
\eeq
\beq
\mathcal{L}_{2}(\phi,\partial \phi)=\sum_{i=1}^{3}\sum_{j=1}^{3}\Omega_{i}\beta_{ij}\Omega_{j}=\frac{(\partial\phi)^{2}}{P_{2}^{2}(\phi)}\sum_{i=1}^{3}\sum_{j=1}^{3}(C_{i}\phi+D_{i})\beta_{ij}(C_{j}\phi+D_{j}) \;\;,\;\;\; \beta_{ij}=\beta_{ji}
\eeq
\beq
X(\phi)=\frac{1}{P_{2}^{2}(\phi)}\Big[ -A_{0}(\lambda_{1}+2\lambda_{2})\phi^{2}+(\lambda_{1}C_{0}-\lambda_{2}B_{0}) \Big]
\eeq
the trace equation (\ref{trace}) takes the form
\beq
\Box_{g}\phi=G(\phi)-X(\phi)(\partial_{\mu}\phi)(\partial^{\mu}\phi)
\eeq
which is the propagation equation for the scalar mode $\phi$. We  see that the situation is similar to the much simpler Theory of the previous subsection, the only difference being that in the generalized case the expressions are much more involved and therefore more freedom to play around with the form of the functions $G(\phi)$ and $X(\phi)$. Again, obviously there exists a parameter space for which $X(\phi)\equiv 0$, $\forall \phi$ and the above equation becomes a generalized Klein-Gordon equation for the potential $V'(\phi)=G(\phi)$, with the prime denoting derivative with respect to $\phi$ here. We conclude, therefore,  that the 11 parameter vacuum Theory (\ref{S}) is equivalent to a generalized Scalar-Tensor Theory.

\subsection{Establishing the Equivalence}
As in the previous section, we now start with the generalized version of (\ref{SactQS}), including all parity even quadratic invariants,
\begin{gather}
S=\frac{1}{2 \kappa}\int d^{n}x \sqrt{-g}\Big[ f(\chi)+f'(\chi)(R-\chi)+b_{1}S_{\alpha\mu\nu}S^{\alpha\mu\nu} +
	b_{2}S_{\alpha\mu\nu}S^{\mu\nu\alpha} +
	b_{3}S_{\mu}S^{\mu}
	\nonumber \\+a_{1}Q_{\alpha\mu\nu}Q^{\alpha\mu\nu} +
	a_{2}Q_{\alpha\mu\nu}Q^{\mu\nu\alpha}  +
	a_{3}Q_{\mu}Q^{\mu}+
	a_{4}q_{\mu}q^{\mu}+
	a_{5}Q_{\mu}q^{\mu} 
	+c_{1}Q_{\alpha\mu\nu}S^{\alpha\mu\nu}+
	c_{2}Q_{\mu}S^{\mu} +
	c_{3}q_{\mu}S^{\mu} \Big]
\end{gather}
After varying with respect to the auxilary field $\chi$ and plugging the result back to the above we obtain the equivalent action
\beq
S=\frac{1}{2\kappa}\int \sqrt{-g}\Big[\Phi R-V(\Phi)+\mathcal{L}_{2}\Big] \label{Seq}
\eeq
Using the post-Riemannian expansion of the scalar curvature, the form of distortion (\ref{distor2}) and the expressions (\ref{omega123}) after some straightforward algebra we find 
\beq
\sqrt{-g} \Phi R=\sqrt{-g}\Phi \tilde{R}-\sqrt{-g}\frac{\omega_{1}(\Phi)}{\Phi}(\partial \Phi)^{2}
\eeq
where
\begin{gather}
-\omega_{1}(\Phi)=(n-1)\frac{\Phi^{2}}{P_{2}(\Phi)^{2}}\Big[ (C_{2}\Phi+D_{2})^{2}+ (C_{3}\Phi+D_{3})^{2}+n(C_{2}\Phi+D_{2})(C_{3}\Phi+D_{3})\Big]+ \nonumber \\
(n-1)\frac{\Phi}{P_{2}(\Phi)}\Big[ (C_{2}-C_{3})\Phi +(D_{2}-D_{3}) \Big]
\end{gather}
Similarly, we compute
\begin{gather}
    \mathcal{L}_{2}=\left( \sum_{i=1}^{3}\sum_{j=1}^{3}(C_{i}\Phi+D_{i})\beta_{ij}(C_{j}\Phi+D_{j}) \right) \frac{(\partial \Phi)^{2}}{P_{2}^{2}(\Phi)} \;\;, \beta_{ij}=\beta_{ji}
\end{gather}
where $\beta_{ij}$ are some linear combinations of the parameters of the quadratic Theory (see appendix B). With the above results the action (\ref{Seq}) takes the form
\beq
S=\frac{1}{2\kappa}\int \sqrt{-g}\Big[\Phi \tilde{R}-V(\Phi)-\frac{\omega(\Phi)}{\Phi}(\partial \Phi)^{2}\Big] \label{EquivS}
\eeq
with the scalar tensor function $\omega(\Phi)$ being given by\footnote{It is interesting to note that for specific choices of the parameters this function can take the form that appears in the so-called 'hybrid' models \cite{Koivisto:2009jn,Capozziello:2015lza,Harko:2018ayt}.}
\begin{gather}
    -\omega(\Phi)=(n-1)\frac{\Phi^{2}}{P_{2}(\Phi)^{2}}\Big[ (C_{2}\Phi+D_{2})^{2}+ (C_{3}\Phi+D_{3})^{2}+n(C_{2}\Phi+D_{2})(C_{3}\Phi+D_{3})\Big]+ \nonumber \\
(n-1)\frac{\Phi}{P_{2}(\Phi)}\Big[ (C_{2}-C_{3})\Phi +(D_{2}-D_{3}) \Big]+\left( \sum_{i=1}^{3}\sum_{j=1}^{3}(C_{i}\Phi+D_{i})\beta_{ij}(C_{j}\Phi+D_{j}) \right) \frac{\Phi}{P_{2}^{2}(\Phi)} \label{om}
\end{gather}
Let us observe that in $\omega(\Phi)$, appear two groups of terms; one having ratios of  (at most) quartic polynomials in $\Phi$ and the other being the ration of two  quadratic (at most) polynomials in $\Phi$. 

Eq. (\ref{EquivS}) establishes the promised equivalence of the Metric-Affine Theory (\ref{S}) to a specific metric and torsionless Scalar-Tensor Theory, the one with a kinetic coupling given by the function (\ref{om}). Some important comments are now in order. Firstly, note that the scalar field potential function $V(\Phi)$ depends only upon the form of the $f(R)$ function and not on the parameters of the quadratic torsion and non-metricity invariants. On the other hand, the kinetic coupling function $\omega(\Phi)$ depends only on the aforementioned 11 dimensionless parameters and is insensitive to the choice of $f(R)$. Therefore the functional form of $f(R)$ governs the shape of the potential, and the quadratic couplings parameters monitor the behavior of the kinetic coupling function.

\subsection{Example: The $f(R)=R+\alpha R^{2}$ case.}

 Unarguably, the most natural choice for the $f(R)$ function is the quadratic Starobinski form $f(R)=R+\alpha R^{2}$ which in the usual metric $f(R)$ Theory is well motivated and widely used for inflationary scenarios. In this case, from the defining relation $\Phi=f'$, it follows that $R=\chi=\frac{(\Phi-1)}{2 \alpha}$ and so $f(\chi(\Phi))=\frac{(\Phi^{2}-1)}{4 \alpha}$ from which we get the usual quadratic potential
 \beq
V(\Phi)=\frac{(\Phi-1)^{2}}{4 \alpha}\label{pot}
 \eeq
It is worth mentioning again, that the form of the $f(R)$ function only fixes the potential, the function $\omega(\Phi)$ depends only on the parameters $a_{i}, b_{j}$ and $c_{k}$ of the quadratic Theory (\ref{S}) and is insensitive to the choice of $f(R)$. Therefore, in this case we get a fixed potential Scalar-Tensor Theory given by
\beq
S=\frac{1}{2\kappa}\int \sqrt{-g}\Big[\Phi \tilde{R}-\frac{(\Phi-1)^{2}}{4 \alpha}-\frac{\omega(\Phi)}{\Phi}(\partial \Phi)^{2}\Big] \label{ST11}
\eeq
and $\omega(\Phi)$ being the same as in (\ref{om}). We conclude therefore that the quadratic $f(R)$ case corresponds to a fixed potential scalar tensor Theory. Consequently the particle spectrum of 
\begin{gather}
	S
	=\frac{1}{2 \kappa}\int d^{4}x \sqrt{-g} \Big[ R+\alpha R^{2}+\
	b_{1}S_{\alpha\mu\nu}S^{\alpha\mu\nu} +
	b_{2}S_{\alpha\mu\nu}S^{\mu\nu\alpha} +
	b_{3}S_{\mu}S^{\mu}+
	a_{1}Q_{\alpha\mu\nu}Q^{\alpha\mu\nu} +
	a_{2}Q_{\alpha\mu\nu}Q^{\mu\nu\alpha} \nonumber \\ +
	a_{3}Q_{\mu}Q^{\mu}+
	a_{4}q_{\mu}q^{\mu}+
	a_{5}Q_{\mu}q^{\mu} 
	+c_{1}Q_{\alpha\mu\nu}S^{\alpha\mu\nu}+
	c_{2}Q_{\mu}S^{\mu} +
	c_{3}q_{\mu}S^{\mu}
	\Big]  \label{SR2}
	\end{gather}
consists of the graviton plus a scalar mode with the potential (\ref{pot}) and kinetic coupling $\omega(\Phi)$ given by (\ref{om}). The non-metric and torsionful MAG Theory \eqref{SR2} is on-shell equivalent to the metric and torsionless Scalar-Tensor Theory \eqref{ST11}.

\subsection{Including Matter}
Of course, it goes without saying that adding matter would drastically change the results obtained here. In general if connection-matter couplings are allowed, the form of the connection field equations changes by the addition of hypermomentum on the right-hand side of (\ref{confe}). Then similarly to the usual Metric-Affine $f(R)$ Theory \cite{sotiriou2010f} the situation depends on the form of these connection couplings and no concrete statement can be made without more information about the couplings. On the other hand, if the matter Lagrangian does not depend explicitly on the independent affine-connection, i.e. in the so-called Palatini case, the results obtained here continue to hold true by merely adding an energy-momentum tensor on the right-hand side of the metric field equations \eqref{mfe}. Indeed, since in this case the connection field equations remain unaltered, the procedure of solving the connection field equations followed here remains essentially  the same, and the resulting on-shell Theory is again  Scalar-Tensor with extra matter apart from the geometrically induced scalar field.

\section{Conclusions}

We have considered an extension of the usual Metric-Affine f(R) Gravity by adding all possible $11$ parity even quadratic torsion and non-metricity invariants. We proved explicitly that the inclusion of the quadratic invariants has a dramatic affect on the dynamics, namely it introduces an additional scalar degree of freedom. The resulting  Theory is equivalent to a metric and torsionless Scalar-Tensor Theory, whose potential and kinetic term coupling depend upon the choice of $f(R)$ and the dimensionless parameters of the original MAG Theory respectively. More precisely, the potential depends solely upon the choice of $f(R)$ alone, whereas the kinetic coupling function depends only on the parameters of the quadratic invariants of the action \eqref{S}. The result is valid for any dimension $n$.

It is remarkable that in this case the scalar field does not arise from some compactification scheme \cite{Faraoni:2004pi} but is rather related to the non-Riemannian nature of spacetime. In our case we see that the breaking of the projective invariance of the $f(R)$ action, by adding the quadratic torsion and non-metricity scalars, has resulted in an extra scalar degree of freedom.  It is also worth noting that this specific class of Scalar-Tensor Theories (with $\Omega(\Phi)$ given by \eqref{om}) acquires a geometrical origin, as given by the mother action \eqref{S}. Reversely, the metric Scalar-Tensor Theories with potential \eqref{om} can be seen as the manifestation of the non-Riemannian degrees of freedom of torsion and non-metricity in this extended Metric-Affine $f(R)$  formulation.

It would be interesting to see how the results of this study are altered/extended by including also the parity violating Hojman (or Holst) term along with the parity odd quadratic torsion and non-metricity invariants.

\section{Acknowledgements}

This work was supported by the Estonian Research Council grant SJD14. I would like to thank Laur Jarv, Margus Saal and Tomi Koivisto  for some useful comments.

\section{Appendix}

\subsection{The coefficients}
 The elements $a_{ij}$ appearing in \eqref{a11}-\eqref{a12} are given in terms of $a_{i}, b_{j}$ and $c_{k}$, as follows
 \beq
a_{11}=4a_{1}-\frac{c_{1}}{2}+4 n a_{3}+2 a_{5}+\frac{(1-n)}{2}c_{2}
 \eeq
 \beq
a_{12}=4 a_{2}+\frac{c_{1}}{2}+2 n a_{5}+4 a_{4}+\frac{(1-n)}{2}c_{3}
 \eeq
 \beq
a_{13}=-2b_{1}+b_{2}+2 c_{1}+2 n c_{2}+2 c_{3}+(1-n)b_{3}
 \eeq
\beq
a_{21}=-\frac{(n-1)}{2}f'+2 a_{2}+\frac{c_{1}}{2}+4 a_{3}+(n+1)a_{5}+\frac{(n-1)}{2}c_{2}
\eeq  
\beq
a_{22}=(n-1)f' +4 a_{1}+2 a_{2}-\frac{c_{1}}{2}+2 (n+1)a_{4}+2 a_{5}+\frac{(n-1)}{2}c_{3}
\eeq
\beq
a_{23}=2(2-n)f'+2 b_{1}-b_{2}-c_{1}+2 c_{2}+(n+1)c_{3}+(n-1)b_{3} \label{a6}
\eeq

\beq
a_{31}=\frac{(n-3)}{2}f'+2 a_{2}+4 a_{3}+(n+1)a_{5} \label{a7}
\eeq
\beq
a_{32}=f'+2\Big(2 a_{1}+a_{2}+a_{5}+(n+1) a_{4}\Big)
\eeq
\beq
a_{33}=2(n-2)f'-c_{1}+2 c_{2}+(n+1)c_{3} \label{a9}
\eeq
Obviously, the $b_{ij}$ and $c_{ij}$ are then given by
\beq
b_{1j}=0\;\;, \;\; c_{1j}=a_{1j}
\eeq
\beq
b_{21}=-\frac{(n-1)}{2}\;\;, \;\; c_{21}=2 a_{2}+\frac{c_{1}}{2}+4 a_{3}+(n+1)a_{5}+\frac{(n-1)}{2}c_{2}
\eeq
\beq
b_{22}=(n-1)\;\;,\;\; c_{22}=4 a_{1}+2 a_{2}-\frac{c_{1}}{2}+2 (n+1)a_{4}+2 a_{5}+\frac{(n-1)}{2}c_{3}
\eeq
\beq
b_{23}=2(2-n)\;\;,\;\;c_{23}=2 b_{1}-b_{2}-c_{1}+2 c_{2}+(n+1)c_{3}+(n-1)b_{3} 
\eeq
\beq
b_{31}=\frac{(n-3)}{2}\;\;, \;\; c_{31}=2 a_{2}+4 a_{3}+(n+1)a_{5} 
\eeq
\beq
b_{32}=1\;\;, \;\;c_{32}=2\Big(2 a_{1}+a_{2}+a_{5}+(n+1) a_{4}\Big)
\eeq
\beq
b_{33}=2(n-2)\;\;, \;\; c_{33}=-c_{1}+2 c_{2}+(n+1)c_{3} 
\eeq
The determinant of the matrix corresponding to the $\{a_{ij}\}$ elements reads
\beq
\det{(A)}=P_{2}(f')=A_{0}(f')^{2}+B_{0}f' +C_{0}
\eeq
where
\beq
A_{0}=a_{13} b_{22} b_{31}+a_{12} b_{23} b_{31}+a_{13} b_{21} b_{32} -a_{11} b_{23} b_{32}-a_{12} b_{21} b_{33} +a_{11} b_{22} b_{33} 
\eeq
\begin{gather}
   B_{0}= a_{13} b_{32} c_{21}-a_{12} b_{33} c_{21} -a_{13} b_{31} c_{22} +a_{11} b_{33} c_{22} +a_{12} b_{31} c_{23} -a_{11} b_{32} c_{23} \nonumber \\ -a_{13} b_{22} c_{31} +a_{12} b_{23} c_{31}
   a_{13} b_{21} c_{32} -a_{11} b_{23} c_{32} +a_{11} b_{22} c_{33} -a_{12} b_{21} c_{33} 
\end{gather}
\beq
C_{0}=-a_{13} c_{22} c_{31}+a_{12} c_{23} c_{31}+a_{13} c_{21} c_{32}-a_{11} c_{23} c_{32}-a_{12} c_{21} c_{33}+a_{11} c_{22} c_{33}
\eeq
The relevant elements of the inverse matrix  $\mathcal{A}^{-1}$ are given by
\beq
P_{2}(f')\tilde{a}_{12} =(a_{13} b_{32} -a_{12} b_{33}) f'+a_{13} c_{32}-a_{12} c_{33}
\eeq
\beq
P_{2}(f')\tilde{a}_{13} =(-a_{13} b_{22} +a_{12} b_{23}) f'-a_{13} c_{22}+a_{12} c_{23}
\eeq
\beq
P_{2}(f')\tilde{a}_{22}=(-a_{13} b_{31} +a_{11} b_{33}) f'-a_{13} c_{31}+a_{11} c_{33}
\eeq
\beq
P_{2}(f')\tilde{a}_{23}=(a_{13} b_{21} -a_{11} b_{23}) f'+a_{13} c_{21}-a_{11} c_{23}
\eeq
\beq
P_{2}(f')\tilde{a}_{32}=(a_{12} b_{31} -a_{11} b_{32}) f'+a_{12} c_{31}-a_{11} c_{32}
\eeq
\beq
P_{2}(f')\tilde{a}_{33}=(-a_{12} b_{21} +a_{11} b_{22}) f'-a_{12} c_{21}+a_{11} c_{22}
\eeq
For completeness let us also report the irrelevant ones
\beq
P_{2}(f')\tilde{a}_{11}=\left(b_{22} f'+c_{22}\right) \left(b_{33} f'+c_{33}\right)-\left(b_{23} f'+c_{23}\right) \left(b_{32} f'+c_{32}\right)
\eeq
\beq
P_{2}(f')\tilde{a}_{21}=\left(b_{23} f'+c_{23}\right) \left(b_{31} f'+c_{31}\right)-\left(b_{21} f'+c_{21}\right) \left(b_{33} f'+c_{33}\right)
\eeq
\beq
P_{2}(f')\tilde{a}_{31}=\left(b_{21} f'+c_{21}\right) \left(b_{32} f'+c_{32}\right)-\left(b_{22} f'+c_{22}\right) \left(b_{31} f'+c_{31}\right)
\eeq
from which we see that indeed only the elements $\tilde{a}_{i1}$ have a numerator quadratic in $f'$, whereas the relevant elements have a numerator that is only linear in $f'$.

\subsection{Post-Riemannian expansions}
Using the post-Riemannian decomposition of the connection, the Ricci tensor and scalar curvature are expanded as
\begin{equation}
R_{\nu\beta}=\tilde{R}_{\nu\beta}+ \tilde{\nabla}_{\mu}N^{\mu}_{\;\;\;\nu\beta}-\tilde{\nabla}_{\beta}N^{(2)}_{\nu}+N^{(2)}_{\lambda}N^{\lambda}_{\;\;\;\nu\beta}-N^{\mu}_{\;\;\;\rho\beta}N^{\rho}_{\;\;\;\nu\mu}
\end{equation}
and 
\begin{equation}
R=\tilde{R}+ \tilde{\nabla}_{\mu}( N^{(3)\mu}-N^{(2)\mu})+ N^{(3)}_{\mu}N^{(2)\mu}-N_{\alpha\mu\nu}N^{\mu\nu\alpha} \label{Recomp}
\end{equation}
respectively. Now,
For the distortion (\ref{distor}), and the above expansions, we readily compute the Ricci tensor
\beq
R_{\mu\nu}=\tilde{R}_{\mu\nu}+\tilde{\nabla}_{\mu}(\Omega_{1}\psi_{\nu})-\tilde{\nabla}_{\nu}\Big( \Omega_{1}+\Omega_{3}+(n-1)\Omega_{2} \Big)\psi_{\mu}+g_{\mu\nu}\tilde{\nabla}_{\alpha}(\Omega_{3}\psi^{\alpha})+\Big[ (n-1)\Omega_{2}^{2}-\Omega_{3}^{2}\Big]\psi_{\mu}\psi_{\nu}+\Omega_{3}\Big[ (n-1)\Omega_{2}+\Omega_{3} \Big]\psi_{\alpha}\psi^{\alpha}g_{\mu\nu} \label{Rmn}
\eeq
with the associated scalar curvature
\beq
R=\tilde{R}+(n-1)\tilde{\nabla}_{\alpha}\Big[(\Omega_{3}-\Omega_{2})\psi^{\alpha} \Big]+(n-1)\Big[ \Omega_{2}^{2}+\Omega_{3}^{2}+n\Omega_{2}\Omega_{3}\Big]\psi_{\mu}\psi^{\mu}
\eeq
For $\psi_{\mu}=\frac{\partial_{\mu}\Phi}{\Phi}$, up to surface terms it holds that
\beq
\sqrt{-g} \Phi R=\sqrt{-g}\Phi \tilde{R}+\frac{(n-1)}{\Phi}\sqrt{-g}\Big[ \Omega_{2}-\Omega_{3}+\Omega_{2}^{2}+\Omega_{3}^{2}+n\Omega_{2}\Omega_{3}\Big](\partial \Phi)^{2}
\eeq

Also note that for the same form of distortion, the Palatini tensor can be easily found to be
\beq
P_{\alpha\mu\nu}=\Big[  (n-1)\Omega_{2}+\Omega_{3}\Big]\psi_{\alpha}g_{\mu\nu}+\Big[\Omega_{2}+(n-1)\Omega_{3} \Big]
\psi_{\mu}g_{\alpha\nu}-(\Omega_{2}+\Omega_{3})\psi_{\nu}g_{\alpha\mu}
\eeq
It is worth stressing the explicit disappearance of $\Omega_{1}$ in both the scalar curvature and Palatini tensor. Of course this is to be expected since the scalar curvature is projective invariant and also the Palatini tensor is insensitive to the projective mode, being the outcome of the $\Gamma$-variation of a projective invariant quantity, namely the aforementioned scalar curvature. As it is clear from (\ref{Rmn}), the same is true for the symmetric part $R_{(\mu\nu)}$ of the Ricci tensor but not for the full tensor. This is so because only the symmetric part of the Ricci tensor is projective invariant.

Furthermore, for the same form of distortion, the various distortion traces and quadratic distortion invariants read
\beq
N_{\mu}^{(1)}=(n \Omega_{1}+\Omega_{2}+\Omega_{3})\psi_{\mu}
\eeq
\beq
N_{\mu}^{(2)}=( \Omega_{1}+n\Omega_{2}+\Omega_{3})\psi_{\mu}
\eeq
\beq
N_{\mu}^{(3)}=( \Omega_{1}+\Omega_{2}+n\Omega_{3})\psi_{\mu}
\eeq
\beq
N_{\alpha\mu\nu}N^{\alpha\mu\nu}=\Big[ n(\Omega_{1}^{2}+\Omega_{2}^{2}+\Omega_{3}^{2})+2(\Omega_{1}\Omega_{2}+\Omega_{2}\Omega_{3}+\Omega_{3}\Omega_{1}) \Big]\psi_{\mu}\psi^{\mu}
\eeq
\beq
N_{\alpha\mu\nu}N^{\mu\nu\alpha}=\Big[ (\Omega_{1}^{2}+\Omega_{2}^{2}+\Omega_{3}^{2})+(n+1)(\Omega_{1}\Omega_{2}+\Omega_{2}\Omega_{3}+\Omega_{3}\Omega_{1}) \Big]\psi_{\mu}\psi^{\mu}
\eeq
\beq
N_{\alpha\mu\nu}N^{\alpha\nu\mu}=\Big[ (\Omega_{1}^{2}+\Omega_{2}^{2}+n \Omega_{3}^{2})+2(n \Omega_{1}\Omega_{2}+\Omega_{2}\Omega_{3}+\Omega_{3}\Omega_{1}) \Big]\psi_{\mu}\psi^{\mu}
\eeq
\beq
N_{\mu\nu\alpha}N^{\alpha\nu\mu}=\Big[ (\Omega_{1}^{2}+n\Omega_{2}^{2}+ \Omega_{3}^{2})+2( \Omega_{1}\Omega_{2}+\Omega_{2}\Omega_{3}+n\Omega_{3}\Omega_{1}) \Big]\psi_{\mu}\psi^{\mu}
\eeq
\beq
N_{\alpha\mu\nu}N^{\mu\alpha\nu}=\Big[ (n\Omega_{1}^{2}+\Omega_{2}^{2}+ \Omega_{3}^{2})+2( \Omega_{1}\Omega_{2}+n\Omega_{2}\Omega_{3}+\Omega_{3}\Omega_{1}) \Big]\psi_{\mu}\psi^{\mu}
\eeq
\beq
N_{\alpha\mu\nu}N^{\nu\alpha\mu}=\Big[ (\Omega_{1}^{2}+\Omega_{2}^{2}+\Omega_{3}^{2})+(n+1)(\Omega_{1}\Omega_{2}+\Omega_{2}\Omega_{3}+\Omega_{3}\Omega_{1}) \Big]\psi_{\mu}\psi^{\mu}
\eeq
with these and given that
\begin{gather}
\mathcal{L}_{2}=\Big( \frac{b_{1}}{2}+ 2 a_{1}-c_{1}\Big) N_{\alpha\mu\nu}N^{\alpha\mu\nu}+	\Big( -\frac{b_{1}}{2}+  a_{2}+c_{1}\Big)N_{\alpha\mu\nu}N^{\alpha\nu\mu} \nonumber \\
+	\Big( \frac{b_{2}}{2}+ 2 a_{2}+c_{1}\Big)N_{\alpha\mu\nu}N^{\mu\nu\alpha}+	\Big( -\frac{b_{2}}{2}+ 2 a_{1}+a_{2}-c_{1}\Big)N_{\mu\nu\alpha}N^{\nu\mu\alpha} \nonumber \\
=\Big( \frac{b_{3}}{4}+4 a_{3}-c_{2} \Big)N^{(1)}_{\mu}N^{(1)}_{\nu}g^{\mu\nu}+\Big( \frac{b_{3}}{4}+a_{4}+\frac{c_{3}}{4} \Big)N^{(2)}_{\mu}N^{(2)}_{\nu}g^{\mu\nu}+a _{4}N^{(3)}_{\mu}N^{(3)}_{\nu}g^{\mu\nu} \nonumber \\
+\Big( 2 a_{5}-\frac{c_{3}}{2}+c_{2}-\frac{b_{3}}{2}\Big) N^{(1)}_{\mu}N^{(2)}_{\nu}g^{\mu\nu}	+\Big( 2 a_{4}+\frac{c_{3}}{2}\Big)N^{(2)}_{\mu}N^{(3)}_{\nu}g^{\mu\nu}+\Big( 2 a_{5}-\frac{c_{3}}{2}\Big) N^{(1)}_{\mu}N^{(3)}_{\nu}g^{\mu\nu}
\end{gather}
we compute
\begin{gather}
\mathcal{L}_{2}=\Big( \frac{b_{1}}{2}+ 2 a_{1}-c_{1}\Big) \Big[ n(\Omega_{1}^{2}+\Omega_{2}^{2}+\Omega_{3}^{2})+2(\Omega_{1}\Omega_{2}+\Omega_{2}\Omega_{3}+\Omega_{3}\Omega_{1}) \Big]\psi_{\mu}\psi^{\mu} \nonumber \\+	\Big( -\frac{b_{1}}{2}+  a_{2}+c_{1}\Big)\Big[ (\Omega_{1}^{2}+\Omega_{2}^{2}+n \Omega_{3}^{2})+2(n \Omega_{1}\Omega_{2}+\Omega_{2}\Omega_{3}+\Omega_{3}\Omega_{1}) \Big]\psi_{\mu}\psi^{\mu} \nonumber \\
+	\Big( \frac{b_{2}}{2}+ 2 a_{2}+c_{1}\Big)\Big[ (\Omega_{1}^{2}+\Omega_{2}^{2}+\Omega_{3}^{2})+(n+1)(\Omega_{1}\Omega_{2}+\Omega_{2}\Omega_{3}+\Omega_{3}\Omega_{1}) \Big]\psi_{\mu}\psi^{\mu}\nonumber \\+	\Big( -\frac{b_{2}}{2}+ 2 a_{1}+a_{2}-c_{1}\Big)\Big[ (n\Omega_{1}^{2}+\Omega_{2}^{2}+ \Omega_{3}^{2})+2( \Omega_{1}\Omega_{2}+n\Omega_{2}\Omega_{3}+\Omega_{3}\Omega_{1}) \Big]\psi_{\mu}\psi^{\mu} \nonumber \\
+\Big( \frac{b_{3}}{4}+4 a_{3}-c_{2} \Big)(n \Omega_{1}+\Omega_{2}+\Omega_{3})^{2}\psi_{\mu}\psi^{\mu}+\Big( \frac{b_{3}}{4}+a_{4}+\frac{c_{3}}{4} \Big)( \Omega_{1}+n\Omega_{2}+\Omega_{3})^{2}\psi_{\mu}\psi^{\mu}\nonumber \\+a _{4}( \Omega_{1}+\Omega_{2}+n\Omega_{3})^{2}\psi_{\mu}\psi^{\mu} \nonumber \\
+\Big( 2 a_{5}-\frac{c_{3}}{2}+c_{2}-\frac{b_{3}}{2}\Big) (n \Omega_{1}+\Omega_{2}+\Omega_{3})(\Omega_{1}+n\Omega_{2}+\Omega_{3})\psi_{\mu}\psi^{\mu}\nonumber \\	+\Big( 2 a_{4}+\frac{c_{3}}{2}\Big)(\Omega_{1}+n\Omega_{2}+\Omega_{3})(\Omega_{1}+\Omega_{2}+n\Omega_{3})\psi_{\mu}\psi^{\mu}\nonumber \\+\Big( 2 a_{5}-\frac{c_{3}}{2}\Big) (n\Omega_{1}+\Omega_{2}+\Omega_{3})(\Omega_{1}+\Omega_{2}+n\Omega_{3})\psi_{\mu}\psi^{\mu}
\end{gather}
or more compactly expressed as
\beq
\mathcal{L}_{2}=(\beta_{11}\Omega_{1}^{2}+\beta_{22}\Omega_{2}^{2}+\beta_{33}\Omega_{3}^{2}+2\beta_{12}\Omega_{1}\Omega_{2}+2\beta_{23}\Omega_{2}\Omega_{3}+2\beta_{31}\Omega_{3}\Omega_{1})\psi_{\mu}\psi^{\mu}
\eeq
with the obvious identifications among the $\beta_{ij}'s$ and $a_{i},b_{j}$ and $c_{k}'s$.

\subsection{Scalar-Tensor Theories}

In order to have a direct comparison of our Theory with the known Scalar-Tensor Theories of Gravity let us include  here the action and the corresponding field equations of generic Scalar-Tensor Theories, in n-dimensions and according to our conventions. The action of a generic  Scalar-Tensor Theory reads
\beq
S[g_{\mu\nu},\Phi,\Psi]=\frac{1}{2 \kappa}\int d^{n}x \sqrt{-g}\Big[ \Phi \tilde{R}-\frac{\omega(\Phi)}{\Phi}(\partial \Phi)^{2}-V(\Phi)+2\kappa \mathcal{L}_{m}(g_{\mu\nu},\Psi)\Big] \label{Jordan}
\eeq
where $\tilde{R}$ is the Riemannian scalar curvature, i.e. Ricci scalar, and $\Psi$ collectively denotes the matter fields whose Lagrangian we denote by $\mathcal{L}_{m}(g_{\mu\nu},\Psi)$. Of course the theory is developed over a Riemannian background where torsion and non-metricity vanish by default. The corresponding field equations that one gets after varying with respect to the scalar field and the metric are the following
\beq
\tilde{R}+\left( \frac{\omega'(\Phi)}{\Phi}-\frac{\omega(\Phi)}{\Phi^{2}}\right)(\partial \Phi)^{2}+\frac{2 \omega}{\Phi}(\Box_{g}\Phi)-V'(\Phi)=0 \label{phife}
\eeq
\beq
\tilde{R}_{\mu\nu}-\frac{\tilde{R}}{2}g_{\mu\nu}=\frac{\kappa T_{\mu\nu}}{\Phi}+\frac{1}{\Phi}(\tilde{\nabla}_{\mu}\tilde{\nabla}_{\mu}-g_{\mu\nu}\Box_{g})\Phi+\frac{\omega(\Phi)}{\Phi^{2}}\left( \partial_{\mu}\Phi \partial_{\nu}\Phi-\frac{1}{2}(\partial \Phi)^{2}g_{\mu\nu}\right)-\frac{V(\Phi)}{2 \Phi}g_{\mu\nu}
\eeq
where $T_{\mu\nu}$ is the energy-momentum tensor corresponding to $\mathcal{L}_{m}$. Taking the trace of the latter metric field equations, it follows that
\beq
\tilde{R}=-\frac{2 \kappa T}{(n-2)\Phi}+2\frac{(n-1)}{(n-2)}\frac{\Box_{g}\Phi}{\Phi}+\frac{\omega(\Phi)}{\Phi^{2}}(\partial \Phi)^{2}+\frac{n}{(n-2)}\frac{V(\Phi)}{\Phi}
\eeq
Using this we can eliminate $\tilde{R}$ from (\ref{phife}) and bring the latter to the more familiar form
\beq
2\left[ \frac{(n-1)}{(n-2)}+\omega \right]\Box_{g}\Phi=\frac{2 \kappa T}{(n-2)}-\omega'(\partial \Phi)^{2}+V'\Phi-\frac{n}{(n-2)}V
\eeq
and for $n=4$ it becomes
\beq
\Box_{g}\Phi=\frac{1}{(3+2 \omega)}\Big[ \kappa T-\omega' (\partial \Phi)^{2}+\Phi V'-2V \Big]
\eeq
Lastly, let us note that one can pass from the Jordan frame representation \eqref{Jordan} to the Einstein frame, where $\tilde{R}$ decouples from $\Phi$, by employing a conformal transformation of the metric:
\beq
g_{\mu\nu} \rightarrow g'_{\mu\nu}=e^{2 \phi}g_{\mu\nu}
\eeq
which induces the transformation laws
\beq
\sqrt{-g'}=e^{n \phi}\sqrt{-g}
\eeq
\beq
\tilde{\Gamma}^{'\lambda}{}_{\mu\nu}=\tilde{\Gamma}^{\lambda}{}_{\mu\nu}+2\delta^{\lambda}_{(\mu}\partial_{\nu)}\phi-(\partial^{\lambda}\phi)g_{\mu\nu}
\eeq
\beq
\tilde{R}'=e^{-2 \phi}\Big( \tilde{R}-2(n-1)\tilde{\nabla}_{\alpha}\tilde{\nabla}^{\alpha}\phi -(n-1)(n-2)(\partial \phi)^{2} \Big)
\eeq

for the volume element, Levi-Civita connection and Riemannian scalar curvature repsectively.

\bibliographystyle{unsrt}
	\bibliography{ref}

\end{document}